\documentclass{Interspeech}

\interspeechcameraready 

\usepackage{ragged2e} 
\usepackage{array}   
\usepackage{enumitem} 
\usepackage{amsmath} 
\usepackage{amsfonts} 
\usepackage{graphicx} 
\usepackage{url} 
\usepackage{siunitx} 

\newcolumntype{L}[1]{>{\RaggedRight\arraybackslash}p{#1}}
\newcolumntype{C}[1]{>{\Centering\arraybackslash}p{#1}}

\title{WARP - Web-Augmented Real-time Program Repairer: A Real-Time Compilation Error Resolution using LLMs and Web-Augmented Synthesis}

\author[affiliation={1}]{Anderson}{de Lima Luiz} 

\affiliation{Technische Hochschule Ingolstadt}{Germany}

\email{anderson.delimaluiz@thi.de}


\begin{document}

\maketitle

\begin{abstract}
Compilation errors represent a significant bottleneck in software development productivity.
This paper introduces WARP (Web-Augmented Real-time Program Repairer), a novel system that leverages Large Language Models (LLMs) and dynamic web-augmented synthesis for real-time resolution of these errors.
WARP actively monitors developer terminals, intelligently detects compilation errors, and synergistically combines the understanding of a fine-tuned Code-LLM with relevant solutions, explanations, and code snippets retrieved from up-to-date web sources like developer forums and official documentation. 
Experimental results on our curated benchmark, CGP (featuring C/C++, Python, and Go errors), demonstrate WARP achieves a superior fix rate (72.5\% Compiles correctly) and higher semantic correctness compared to baseline LLM-only approaches and traditional IDE quick-fixes. Key technical challenges in achieving high-accuracy synthesis from noisy web data.
\end{abstract}

\section{Introduction}
\label{sec:introduction}
The iterative nature of software development is frequently punctuated by compilation errors, whose diagnosis and resolution consume disproportionate developer time and cognitive resources \cite{Bhatia19DeepDelta, Parnin11AreAutomated}.
Existing Integrated Development Environments (IDEs), while offering some assistance, often fall short with complex, context-dependent, or novel errors, frequently providing generic suggestions \cite{Becker16Effective}.  
This paper introduces WARP (Web-Augmented Real-time Program-repairer), a system advancing automated compilation error resolution by synergizing Large Language Models (LLMs) \cite{Huynh25SurveyLLMCode, Chen21Codex, Touvron23Llama2} with vast, dynamic web knowledge.

The primary contributions are: 
(i) a novel, comprehensive architecture for WARP; 
(ii) specific techniques for AST-based contextual feature extraction, prompt engineering for code-centric LLMs, and effective LLM-RAG synergy for compilation errors; and 
(iii) empirical validation of WARP's efficacy on ComErrFix-CGP-v1.1, a new benchmark of C/C++, Python, and Go errors, showing significant improvements over baseline approaches.

The remainder of this paper is structured as follows. Section \ref{sec:related_work} reviews related work. Section \ref{sec:warp_system} details the architecture and methodology of WARP. Section \ref{sec:evaluation} presents our experimental setup, benchmark, and results. Section \ref{sec:discussion} discusses challenges and future research directions. Finally, Section \ref{sec:conclusion} concludes the paper.

\begin{figure}[t]
  \centering
  \includegraphics[width=\linewidth]{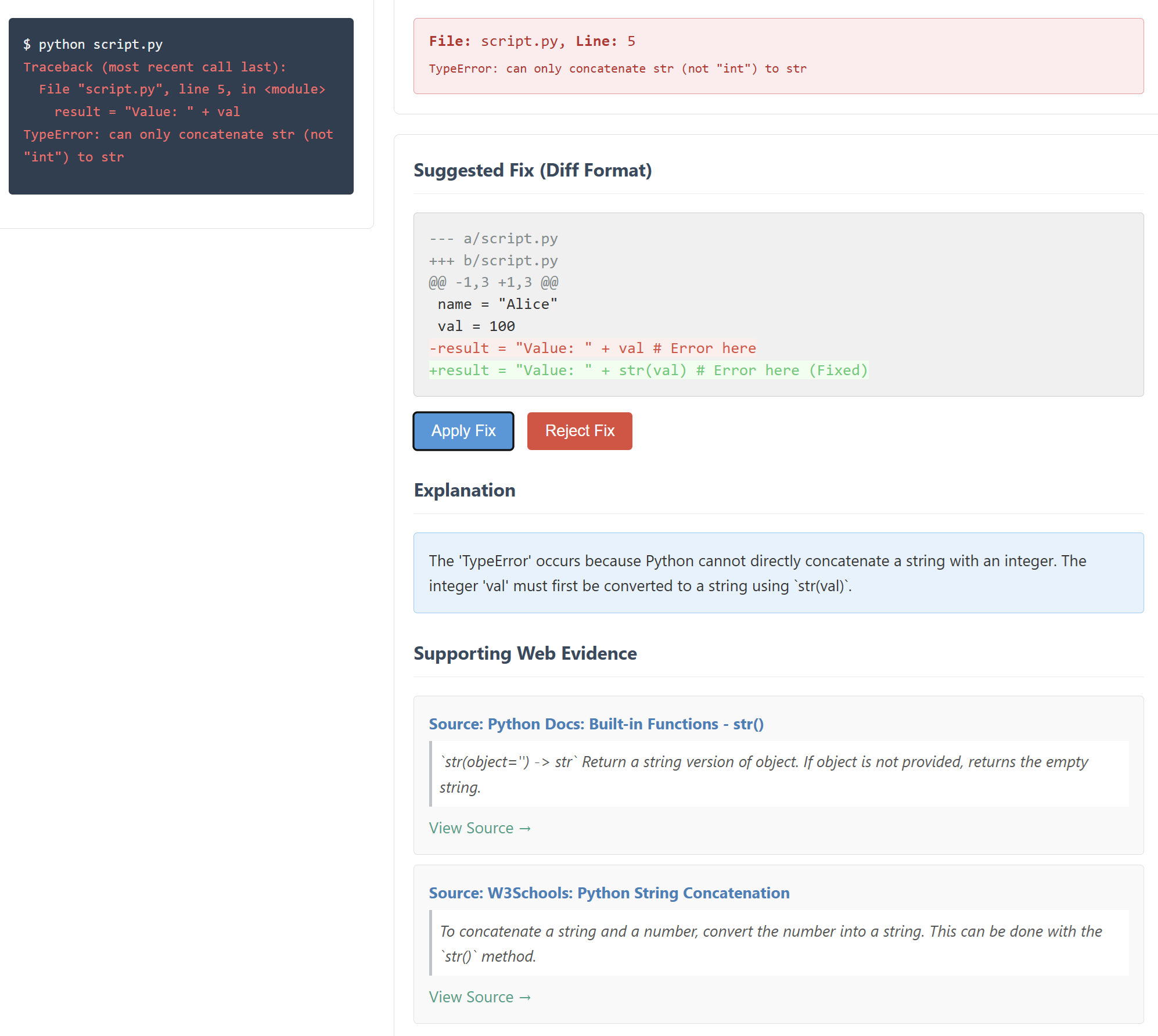}
  \caption{Interface of the WARP system, displaying a detected error, the LLM-generated explanation, suggested fix (diff), and supporting web evidence. This interface aims for clarity and actionable insights for the developer.}
  \label{fig:platform_screenshot}
\end{figure}

\section{Related Work}
\label{sec:related_work}
WARP intersects with real-time error analysis, Automated Program Repair (APR), LLMs applied to code, and Information Retrieval (IR) for software engineering.
Contemporary IDEs integrate static analysis tools for real-time feedback \cite{Ayewah08FindBugs}, but suggestions are often limited to simple syntactic errors, lacking deep semantic understanding or awareness of broader community solutions.

APR \cite{LeGoues19APRSurvey} has a rich history, including search-based, template-based, and constraint-based methods.
LLMs have spurred learning-based APR. DeepDelta \cite{Bhatia19DeepDelta} used NMT for specific Java compilation errors, predating modern LLMs. 
More recent tools like RustAssistant \cite{Deligiannis24RustAssistant} leverage LLMs like GPT-4 \cite{OpenAI23GPT4} for Rust compilation errors via iterative compiler feedback; however, it is language-specific and lacks dynamic web search.
Other systems like SRepair \cite{Xia23CanLLM} and RepairAgent \cite{Xia24RepairAgent} use LLMs for runtime bugs from benchmarks like Defects4J \cite{Just14Defects4J}, typically operating offline and targeting different bug categories than WARP's focus on real-time compilation error resolution augmented by web knowledge.

Foundational LLMs like OpenAI's Codex \cite{Chen21Codex} and GPT series \cite{OpenAI23GPT4}, and Meta's Llama family \cite{Touvron23Llama2, Roziere23CodeLlama}, excel at code tasks but can "hallucinate" \cite{Ji23HallucinationSurvey} and may lack knowledge of recent API changes or specific error contexts post-training. 
Retrieval Augmented Generation (RAG) \cite{Lewis20RAG} mitigates these issues by grounding LLM responses in external information. 
Applying RAG effectively to diverse, noisy developer-focused web data for real-time code repair, as WARP does, poses challenges in query formulation, source selection, information extraction, low-latency processing, and synthesis of potentially conflicting information \cite{Gao23RAGSurvey, Wang24SpeculativeRAG}. WARP's architecture is designed to address these.

\section{The WARP System: Architecture and Methodology}
\label{sec:warp_system}
\begin{figure}[htbp]
  \centering
   \includegraphics[width=\linewidth]{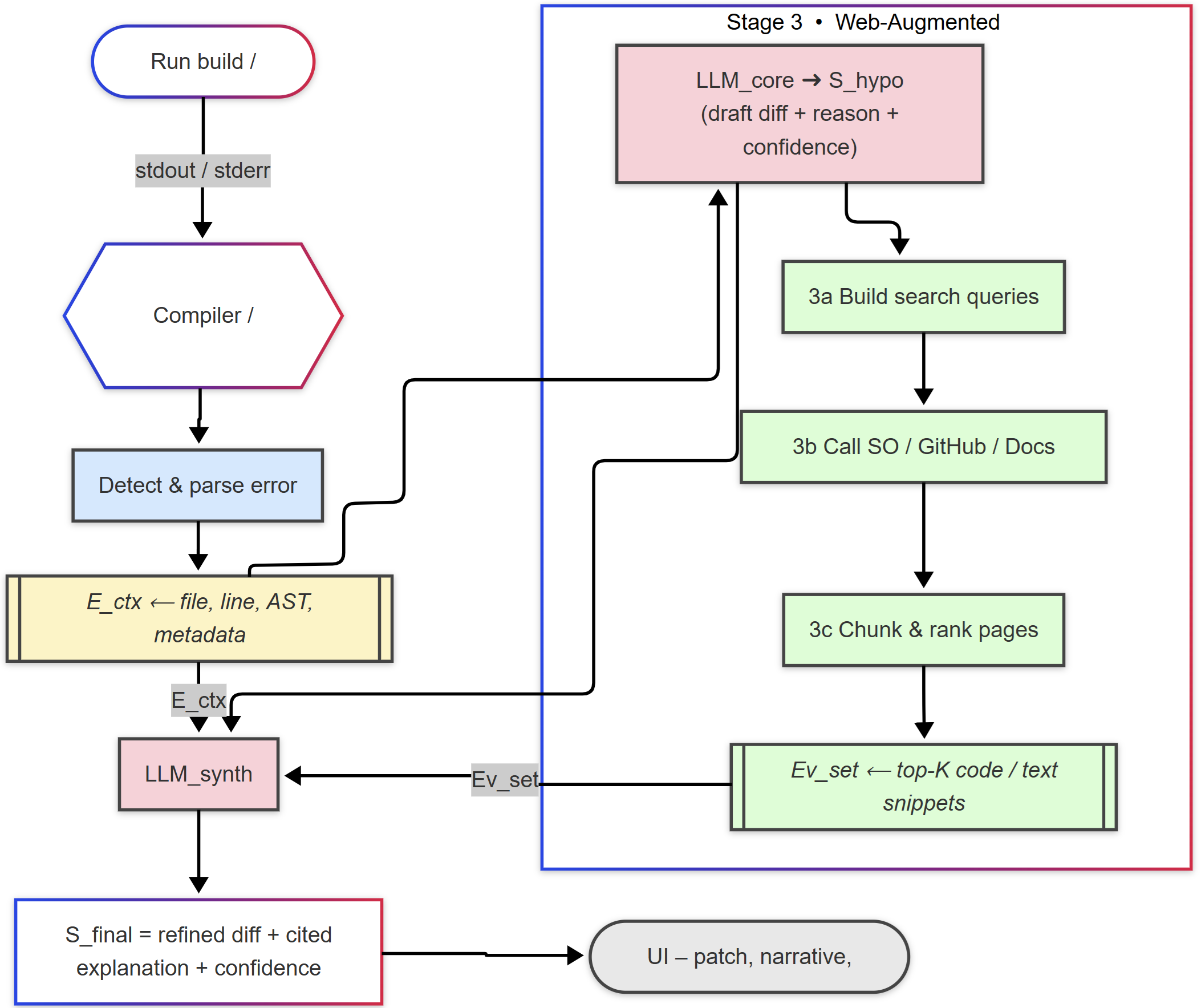}
  \caption{High-level architecture of the WARP system, illustrating information flow from real-time error capture, through contextual feature extraction, initial LLM analysis, web augmentation via the WAR-Module, and final LLM-driven synthesis, before presenting the solution to the developer.}
  \label{fig:system_diagram}
\end{figure}

WARP's operational pipeline, depicted conceptually in Figure \ref{fig:system_diagram}, initiates with real-time capture of compiler outputs.
Upon error detection, it formulates a rich contextual prompt incorporating the error message, Abstract Syntax Tree (AST)-derived code snippets, and project metadata.
This context is fed to a fine-tuned CodeLlama-70B variant \cite{Roziere23CodeLlama} for initial analysis and a candidate fix. 
Concurrently, WARP's specialized Retrieval Augmented Generation (RAG) module \cite{Lewis20RAG, Singh25AgenticRAGSurvey} queries web sources like Stack Overflow, GitHub Issues \cite{Vasilescu13StackOverflowGitHub}, and official documentation to retrieve relevant code examples, explanations, and solutions. 
In the final stage, the LLM synthesizes its pre-trained knowledge with this retrieved web intelligence to generate ranked potential fixes (as diffs), detailed explanations, and references to supporting web evidence, presented via the interface shown in Figure \ref{fig:platform_screenshot}.
The system synergizes deep LLM analytics with dynamic web knowledge for real-time, low-latency operation.

\subsection{Real-time Error Ingestion and Contextual Feature Extraction}
WARP monitors `stdout` and `stderr` from compilation commands (e.g., `gcc`, `python script.py`) or integrates with Language Server Protocol (LSP) implementations. On detecting error signatures, the Ingestion module parses compiler/interpreter output to extract a structured error context, $E_{ctx}$.
This context includes:
\begin{itemize}[noitemsep,topsep=0pt,leftmargin=*]
    \item \textbf{Canonicalized Error Identifier ($E_{id}$)}: Standardized error type (e.g., \texttt{C\_SEMICOLON\_EXPECTED}).
    \item \textbf{Tokenized Error Message ($M_{tok}$)}: Primary error message, tokenized.
    \item \textbf{File Path ($F_{path}$)}: Source file path.
    \item \textbf{Line Number ($L_{num}$)}: Error line number.
    \item \textbf{AST-based Code Context ($C_{AST}$)}: AST nodes around $L_{num} \pm k$ lines, enclosing function, or block, generated using a parser (e.g., `tree-sitter`). Provides richer structural information than raw text.
    \item \textbf{Programming Language ($L_{lang}$)}: Inferred from file extension or command.
    \item \textbf{Project-Specific Metadata ($P_{meta}$)}: Dependencies, library versions, compiler options from build files (e.g., `pom.xml`, `requirements.txt`) or flags.
\end{itemize}
This $E_{ctx} = f_{extract}(\text{raw\_compiler\_output}, \text{source\_file\_content}, \text{project\_config})$ is input for subsequent stages.

\subsection{LLM-Internal Hypothesis Generation}
The structured $E_{ctx}$ is formatted into an initial prompt, $P_{hypo}$, processed by a core LLM, $\text{LLM}_{core}$ (e.g., a fine-tuned CodeLlama-70B variant or similar). 
This $\text{LLM}_{core}$ is fine-tuned on a large dataset of erroneous code snippets, error messages, metadata, corrected code diffs, explanations, and error categories to maximize correction accuracy and enhance explanation generation in diff format.

$P_{hypo}$ guides $\text{LLM}_{core}$ for analysis and candidate fix generation. A template:
\begin{quote}
\small
\texttt{Task: Analyze and resolve a compilation error.} \\
\texttt{Language: \$\{L\_lang\}} \\
\texttt{Error Type: \$\{E\_id\}} \\
\texttt{Error Message: "\$\{M\_tok\}"} \\
\texttt{File: \$\{F\_path\} at Line: \$\{L\_num\}} \\
\texttt{AST Context Snippet: \$\{C\_AST\}} \\
\texttt{Relevant Project Metadata: \$\{P\_meta\}} \\
\texttt{Instruction: Provide a concise textual explanation of the root cause. Then, provide a suggested code modification in 'diff -u' format. Estimate confidence.}
\end{quote}
$\text{LLM}_{core}$ generates $S_{hypo} = (\text{Fix}_{hypo}, \text{Expl}_{hypo}, \text{Conf}_{hypo}) = \text{LLM}_{core}(P_{hypo})$, containing the proposed diff, explanation, and an initial confidence score (e.g., from token log-probabilities or a confidence head).

\subsection{Web-Augmented Retrieval (WAR) Module}
To complement $\text{LLM}_{core}$'s knowledge, the WAR module, often invoked in parallel or post-$S_{hypo}$ generation, performs:
\begin{enumerate}[leftmargin=*,label=\arabic*),nosep,topsep=0pt,partopsep=0pt]
    \item \textbf{Dynamic and Contextual Query Formulation}: Generates diverse search queries $\{Q_1, ..., Q_N\}$ using $M_{tok}$, $L_{lang}$, $E_{id}$, keywords from $\text{Expl}_{hypo}$ (if available), and $P_{meta}$ (e.g., library versions). So, $Q_i = f_{qgen}(M_{tok}, L_{lang}, E_{id}, \text{keywords}(\text{Expl}_{hypo}), P_{meta})$.
    \item \textbf{Targeted Multi-Source Search}: Dispatches queries to:
        \begin{itemize}[noitemsep,topsep=0pt,leftmargin=2em]
            \item \textbf{Stack Overflow API}: For questions tagged with $L_{lang}$ and matching $Q_i$.
            \item \textbf{GitHub Issues Search API}: Within relevant repositories or globally.
            \item \textbf{General Web Search API} (e.g., Google, Bing): Prioritizing official documentation and reputable blogs.
        \end{itemize}
    \item \textbf{Document Snippet Retrieval, Chunking, and Relevance Ranking}: Fetches top $K$ documents per query, segments them into snippets $\{ev_1, ..., ev_M\}$, and ranks each $ev_j$ using $Score(ev_j | E_{ctx})$ based on semantic similarity, source reputation, recency, keyword relevance, etc. The top $M'$ non-redundant evidence snippets, $Ev_{set}$, are selected.
\end{enumerate}

\subsection{LLM-based Synthesis and Solution Refinement}
An $\text{LLM}_{synth}$ model (which could be $\text{LLM}_{core}$ or a specialized variant) synthesizes $S_{hypo}$ with $Ev_{set}$.
A comprehensive synthesis prompt, $P_{synth}$, includes $E_{ctx}$, $S_{hypo}$, and $Ev_{set}$. The LLM is tasked to:
\begin{quote}
\small
\texttt{Given original error context: \$\{E\_ctx\}, initial hypothesis: Fix='\$\{Fix\_hypo\}', Explanation='\$\{Expl\_hypo\}'.} \\
\texttt{Review web evidence: \$\{Ev\_set\}.} \\
\texttt{Synthesize to:}
\texttt{1. Confirm/refine explanation, citing evidence.}
\texttt{2. Confirm/refine code fix (diff). Prioritize better web-suggested approaches, explaining rationale.}
\texttt{3. Reconcile/highlight conflicting web evidence.}
\texttt{4. Provide overall confidence for the final solution.}
\texttt{Output: (1) refined diff, (2) synthesized explanation with citations, (3) utilized evidence IDs, (4) final confidence.}
\end{quote}
The output $S_{final} = (\text{Fix}_{final}, \text{Expl}_{final}, \text{Cit}_{web}, \text{Conf}_{final})$ is generated. $\text{Conf}_{final}$ is recalibrated by $\text{LLM}_{synth}$ based on coherence with web evidence. The top-ranked $S_{final}$ is presented to the user.

\section{Experimental Evaluation}
\label{sec:evaluation}
We assessed WARP on **ComErrFix-CGP-v1.1**, our new benchmark of 2000 unique compilation errors: C/C++ (700), Python (700), Go (600), from GitHub, academic assignments, and Stack Overflow. 
Each instance includes: (i) erroneous code (10-50 lines); (ii) error message(s); (iii) project context (dependencies, flags); (iv) canonical ground-truth diff; (v) human-authored explanation; (vi) up to three human-verified relevant web URLs.

\subsection{Baselines and Metrics}
Compared WARP against:
\begin{enumerate}[noitemsep,topsep=0pt,leftmargin=*]
    \item \textbf{IDE-Std}: Simulated typical IDE "quick-fixes" using pattern-matching for simple errors.
    \item \textbf{$\text{LLM}_{core}$-Only}: WARP's fine-tuned CodeLlama-70B variant without the WAR module, processing $P_{hypo}$ from $E_{ctx}$.
    \item \textbf{RAG-BasicTFIDF}: Simpler RAG using $P_{hypo}$, TF-IDF/BM25 retrieval from a static Stack Overflow dump, concatenating the top passage to $P_{hypo}$ for single-pass $\text{LLM}_{core}$ generation. Lacks WARP's advanced query, multi-source search, ranking, and synthesis.
\end{enumerate}
Primary metrics:
\begin{itemize}[noitemsep,topsep=0pt,leftmargin=*]
    \item \textbf{Fix Rate (ExactMatch)}: \% top-1 diffs matching ground-truth.
    \item \textbf{Fix Rate (CompilesCorrectly)}: \% top-1 diffs that compile/run correctly and are semantically correct (assessed by unit tests for 30\% of benchmark, manual review for others).
    \item \textbf{Explanation Quality (BLEU-4 \& ROUGE-L)}: Vs. human-authored explanations.
    \item \textbf{Evidence Relevance (NDCG@3)}: For RAG-BasicTFIDF and WARP-Full, ranking quality vs. human-verified URLs.
    \item \textbf{Mean Reciprocal Rank (MRR) for Fixes}: Ranking quality of first correct fix if multiple suggestions. (Primarily top-1 reported).
    \item \textbf{Latency (E2E)}: Average wall-clock time (s) from error ingestion to solution presentation.
\end{itemize}

\subsection{Results and Discussion}
Key results are in Table \ref{tab:results_detailed}. 
WARP-Full significantly improved over baselines in fix accuracy and explanation quality, achieving a 72.5\% CompilesCorrectly fix rate (vs. 51.2\% for $\text{LLM}_{core}$-Only, 59.8\% for RAG-BasicTFIDF). Semantic Correctness (SC) within CompilesCorrectly fixes was also higher for WARP-Full (94

\begin{table}[thbp]
  \caption{Performance Comparison on ComErrFix-CGP-v1.1. SC = Semantic Correctness for compiling fixes. All values are percentages (\%) unless specified as (s) for seconds. Best performance in bold.}
  \label{tab:results_detailed}
  \centering
  \resizebox{\columnwidth}{!}{%
  \begin{tabular}{lccccc}
    \toprule
    \textbf{System} & \textbf{ExactMatch} & \textbf{Compiles} & \textbf{BLEU-4} & \textbf{NDCG@3} & \textbf{Latency} \\
     & \textbf{Fix (\%)} & \textbf{Correctly (\%)} & \textbf{(Expl.)} & \textbf{(Evidence)} & \textbf{(s)} \\
    \midrule
    IDE-Std & 6.1 & 7.5 (SC: 98) & N/A & N/A & \textbf{0.2} \\
    $\text{LLM}_{core}$-Only & 38.5 & 51.2 (SC: 88) & 0.32 & N/A & 2.1 \\
    RAG-BasicTFIDF & 45.3 & 59.8 (SC: 89) & 0.35 & 0.51 & 3.9 \\
    \textbf{WARP-Full} & \textbf{60.2} & \textbf{72.5 (SC: 94)} & \textbf{0.45} & \textbf{0.78} & 4.2 \\
    \bottomrule
  \end{tabular}%
  }
\end{table}

WARP-Full's superior Explanation Quality (BLEU-4: 0.45) and Evidence Relevance (NDCG@3: 0.78) highlight the WAR module's effectiveness in sourcing pertinent information and $\text{LLM}_{synth}$'s ability to integrate it. IDE-Std's low latency is offset by its limited applicability. WARP-Full's E2E latency of \SI{4.2}{\second}, while higher than $\text{LLM}_{core}$-Only, is acceptable for interactive real-time assistance.

Qualitative analysis showed WARP's strength in resolving errors requiring knowledge beyond LLM training data (e.g., recent API changes, undocumented behaviors). For instance, with post-training breaking changes in a Go networking library, $\text{LLM}_{core}$-Only failed or suggested outdated patterns. WARP retrieved relevant GitHub discussions and updated documentation, enabling $\text{LLM}_{synth}$ to propose the correct fix with citations.

\subsection{Ablation Study: Impact of WAR Module and Synthesis}
Removing the WAR module ($\text{LLM}_{core}$-Only) dropped ExactMatch Fix Rate by ~21.7\% and CompilesCorrectly by ~21.3\% vs. WARP-Full. RAG-BasicTFIDF performed significantly worse than WARP-Full (deficit of ~14.9\% ExactMatch, ~12.7

\section{Discussion, Challenges, and Future Work}
\label{sec:discussion}
WARP's architecture, combining a fine-tuned LLM with dynamic web retrieval and synthesis, significantly advances real-time compilation error resolution. Persistent challenges and future work include:

\textbf{Latency Optimization}: While \SI{4.2}{\second} is viable, further reduction is desirable. Work includes model quantization, knowledge distillation, caching, and speculative execution.

\textbf{Robustness to Web Noise}: The internet contains outdated or incorrect information. Enhancing WAR module's quality assessment (source credibility, temporal decay) and $\text{LLM}_{synth}$'s handling of conflicting evidence is crucial.

\textbf{Handling Complex/Cascading Errors}: WARP excels at single errors. Future work will explore diagnosing interdependencies in multiple errors and proposing sequential/batched fixes, possibly with interactive clarification.

\textbf{Deeper Project Context Integration and Personalization}: Current project metadata ($P_{meta}$) usage is local. Future work aims to index/query the entire codebase for globally-aware reasoning and personalize suggestions based on developer/team history.

\textbf{Ensuring Security/Reliability of Fixes}: Suggested fixes could introduce vulnerabilities. Integrating SAST tools and extensive semantic validation (beyond basic compilation/unit tests) as post-synthesis steps is planned.

\textbf{User Trust, Explainability, and Control}: Trust requires accuracy and transparent explanations. Improving $\text{LLM}_{synth}$'s reasoning clarity and providing user control over sources/fix types are key.

\section{Conclusion}
\label{sec:conclusion}
This paper introduced WARP, a novel system for real-time, web-augmented compilation error resolution. By integrating a fine-tuned Code-LLM with a specialized Web-Augmented Retrieval (WAR) module and a sophisticated synthesis process, WARP delivers actionable, context-relevant, evidence-backed fixes. We detailed its methodology, including AST-based contextual feature extraction, dynamic query generation, multi-faceted evidence ranking, and LLM-driven synthesis.

Our experimental evaluation on the ComErrFix-CGP-v1.1 benchmark (C/C++, Python, Go) showed WARP significantly outperforms baselines, achieving 72.5\% for semantically correct, compiling fixes, with superior explanation quality and evidence relevance. This highlights the value of targeted, up-to-date web knowledge and intelligent synthesis.

While challenges in latency, web noise robustness, complex error handling, and fix security remain, WARP provides a strong foundation for next-generation AI-powered developer tools. Future work will address these, aiming for deeper project context integration, personalization, and enhanced explainability, to make debugging compilation errors more efficient.

\ifinterspeechfinal 
\else
\fi
\bibliographystyle{IEEEtran}
\bibliography{submission.bib} 

\end{document}